\documentclass[twocolumn,showpacs,preprintnumbers,amsmath,amssymb]{revtex4}
\usepackage{graphicx,subfigure,amssymb,amsmath,enumerate}

\newcommand{\Eq}[1]{Eq.\ (\ref{#1})} 
\newcommand{\Fig}[1]{ Fig.\ \ref{#1}}
\newcommand{\Tab}[1]{ Tab.\ \ref{#1}}
\newcommand{\Sect}[1]{ Sec. \ref{#1}}

\newcommand{\R}       {\mathbb{R}}

\newcommand{\Slocal}{\mathsf{S}^{\text{local}}_0}
\newcommand{\Sglobal}{\mathsf{S}^{\text{global}}_0}
\newcommand{\argmax}       {\textrm{argmax}}
\newcommand{\ext}{\beta^{\text{ext}}}
\newcommand{\open}{\alpha^{\text{open}}}

\begin{document}
\title{Finite-temperature local protein sequence alignment: \\
  percolation  and free-energy distribution}
\author{S Wolfsheimer
  \email{stefan.wolfsheimer@parisdescartes.fr}
}
\altaffiliation{
Institut f\"ur Physik,\\
Carl von Ossietzky Universit\"at Oldenburg,\\
D-26111 Oldenburg, Germany  }

\affiliation{MAP5, U.F.R. de Mat\'ematiques et Informatique,\\
Universit\'e Paris Descartes,\\
45 rue des Saint-P\`eres\\
F-75270 Paris Cedex 06, France}

\author{O Melchert
  \email{melchert@theorie.physik.uni-oldenburg.de}
}
\affiliation{
Institut f\"ur Physik,\\
Carl von Ossietzky Universit\"at Oldenburg,\\
D-26111 Oldenburg, Germany  }

\author{AK Hartmann %
  \email{alexander.hartmann@uni-oldenburg.de}
}
\affiliation{
Institut f\"ur Physik,\\
Carl von Ossietzky Universit\"at Oldenburg,\\
D-26111 Oldenburg, Germany }


\date{\today}

\begin{abstract}
Sequence alignment is a tool in bioinformatics that is used to find
homological relationships in large molecular databases.  It can be
mapped on the physical model of directed polymers in random media.  We
consider the finite-temperature version of local sequence alignment
for proteins and study the transition between the linear phase and the
biologically relevant logarithmic phase, where the free-energy grows
linearly or logarithmically with the sequence length.  By means of
numerical simulations and finite-size scaling analysis we determine
the phase diagram in the plane that is spanned by the gap costs and
the temperature. We use the most frequently used parameter set for
protein alignment.  The critical exponents that describe the parameter
driven transition are found to be explicitly temperature dependent.

Furthermore, we study the shape of the (free-) energy distribution
close to the transition by rare-event simulations down to
probabilities of the order $10^{-64}$.  It is well known that, in the
logarithmic region, the optimal score distribution ($T=0$) is
described by a modified Gumbel distribution. We confirm that this also
applies for the free-energy distribution ($T>0$).  However, in the
linear phase, the distribution crosses over to a modified Gaussian
distribution.
\end{abstract}

\maketitle

%
%
\section{Introduction}
Biological sequence analysis is an interdisciplinary scientific field
which uses concepts from statistics, computer science and molecular
biology.  Some approaches used in the context of sequence
analysis are, from a conceptional point of view, related to models in
statistical mechanics of disordered systems.  One of the most
fundamental tools in the area of sequence analysis is \emph{sequence
alignment} (see for example \cite{Durbin1998,Clote2005}).  It is used
to quantify similarities between two (or more) biological sequences,
like DNA, proteins or RNA.  Modern search tools for large databases,
like BLAST \cite{Altschul1990} or FASTA \cite{Pearson1988}, heavily
rely on sequence alignment algorithms.

In this article we consider algorithms for pairwise local protein
alignment which aims at finding ``conserved'' regions of two input
protein sequences.  The most prominent example is the Smith-Waterman
algorithm \cite{Smith1981}.  The algorithm finds optimal alignments
(OA) according to an objective function.  Each alignment is assigned a
score which is maximal for optimal alignments.  The optimal alignment
score serves as a scalar measure of similarity of the input sequences.
Since alignments have a geometrical interpretation as directed paths
\cite{Zhang1995}, the problem of finding an optimal alignment is
directly related to the ground state of directed paths in random media
\cite{Huse1985,Kardar1987,Mezard1990,Fisher1991,Kardar1994} (DPRM) in
$1+1$ dimensions.  From this point of view, the alignment score
corresponds to the negative energy and the optimal alignment to the
ground state of the system.

However, in some cases optimal alignments are not desirable and one is
interested in ensembles of probabilistic alignments. This is
particularly the case when one wishes to compare so called weak
homologs, i.e. sequences that are related on a relatively long
evolutionary time scale.  In the literature some examples can be
found, where probabilistic alignments clearly outperform optimal
alignments \cite{Muckstein2002,Jaroszewski2002,Koike2004,Koike2007}.
From the physical perspective, a natural generalization to
probabilistic alignments can be achieved by introducing a temperature
and considering canonical ensembles of alignments for each pair of
input sequences instead of the ground state alone
\cite{Miyazawa1995,Kschischo2000,Muckstein2002}.  The finite
temperature approach provides also interesting applications when one wishes
to assess the reliability of alignments by so called posterior
probabilities \cite{Miyazawa1995,Durbin1998}.

For both approaches, for OA and for finite-temperature alignments
(FTA), the choice of the algorithmic parameters remains ambiguous.  In
particular, the choice of the so called gap-costs (see below) requires
some heuristic experimentation.  Interestingly, this question can be
approached by the theory of critical phenomena. The study of sequence
alignment from that perspective yields interesting results that have
improved the optimal choice of parameters of sequence alignment
\cite{Olsen1999,Drasdo2000,Kschischo2000}.  The
\emph{linear-logarithmic phase transition}
\cite{Waterman1987,Arratia1994,Bundschuh2000} is the most important
aspect regarding this issue.  The name stems from the fact that there
is a continuous, parameter-driven transition between phases where the
average score grows linearly or logarithmically with sequence length,
respectively.  There is much empirical evidence that the optimal
choice of scoring parameters is close to the phase boundary on the
logarithmic side \cite{Vingron1994,Hwa1998}.  The underlying reason is
that the transition is driven by the balance between the \emph{score
matrix} that measures the similarity between letters of the underlying
alphabet (i.e.\ amino acids in the case of protein alignment) and the
gap costs.  The latter ones control how strong insertions or deletions
of subsequences are to be penalized.  Hence, one would like to
identify similar regions, which means to try to avoid gaps, giving
them a penalty.  On the other hand, one would like to ignore small
local evolutionary changes to the sequences, which means one should
not make the gap penalty too strong.  This leads to an optimum choice
of the gap penalty parameters at ``intermediate values''.

At $T=0$, i.e.\ for OA, Hwa and L{\"a}ssig have studied the transition
by looking at the dynamic growth of the local (and global) score when
advancing in the search space \cite{Hwa1998}.  Later, the critical
values were studied analytically by a self-consistent equation
\cite{Bundschuh2000} or numerically by a finite-size scaling analysis
\cite{Sardiu2005}.  Both studies rely on a simple scoring model with a
single mismatch parameter.  In the latter procedure the problem was
approached by considering the linear-logarithmic phase transition as a
percolation phenomenon \cite{Stauffer}.

The aim of our study is to go beyond the models that have been
considered so far. In particular, we studied the most widely used
protein alignment model, i.e.~local alignment with the scoring matrix
{\tt blosum62} \cite{Heinkoff1992} and affine gap costs (see below),
where the linear-logarithmic phase transition is of actual relevance
to the database queries or alignment analysis of protein sequences.

We considered the geometrical interpretation of alignments and studied
numerically the percolation properties of OA and FTA.  This allowed us
to determine critical exponents that describe the parameter driven
linear-logarithmic phase transition.  Furthermore we determined the
phase diagram in the plane that is spanned by the temperature and the
gap costs.  Finally, we studied the distribution of the optimal score
and the free energy close to the transitions.

In the following section we review the model and algorithms to compute
the partition functions and methods to sample alignments from the
canonical ensemble.  The main results for different observables and
the (free-) energy distributions are presented in \Sect{sect:results},
followed by a discussion in \Sect{sect:discussion}.

%
%
\section{Partition function calculation and sampling}
\label{sect:partition-function}
An alignment relates letters from one sequence ${\bf a}=a_1\ldots a_L
\in \Sigma^L$ to a second one ${\bf b}=b_1 \ldots b_M \in \Sigma^M$
where $\Sigma$ denotes the underlying alphabet.  Here, we consider
protein sequences wherein $\Sigma$ is given by the 20 letter amino
acid alphabet.  Given the pair ${\bf a}$ and ${\bf b}$, an alignment
$\mathcal{A}$ is an ordered set of pairings $\left\{ (i_1, j_1),
\ldots, (i_{N_m}, j_{N_m}) \right\}$ with $1 \leq i_k < i_{k+1} \leq
L$, $1 \leq j_k < j_{k+1} \leq M$.  If $a_{i_k} = b_{j_k}$ the pair
$(i_k,j_k)$ is called a match otherwise a mismatch. Consequently, we
will refer to $N_m$ as the number of matches plus mismatches.  The
state space of all alignments of ${\bf a}$ and ${\bf b}$ shall be
written as $\mathcal{\chi}_{{\bf a},{\bf b}}$.

When comparing sequences, one has to account for so called insertions
or deletions of subsequences that occur in evolutionary processes.
Regarding alignments, these processes are represented by gaps, which
are defined as follows.  If $i_{k+1} = i_k+1$ and $j_{k+1} = j_k+1+l$
with $l > 0$ and $(i_k,j_k),(i_{k+1},j_{k+1}) \in \mathcal{A}$, then
${\bf b}$ is said to contain a \emph{gap of length $l$ between $j_k$
and $j_{k+1}$} and likewise for the sequence ${\bf a}$.  If $j_1 = l+1
\ge 2$, then ${\bf b}$ is said to have a \emph{gap of length $l$ at
the begin}, if $j_N = M-l < M$, then ${\bf b}$ has a \emph{gap of
length $l$ at the end} and likewise for the sequence ${\bf a}$.

For the comparison of sequences its relevant to give a measure for the
similarity or the degree of conservation between the sequences or
regions of the sequences under consideration.  The classical way to
accomplish this is to assign a \emph{score} for each alignment via an
\emph{objective function} $\mathcal{S} : \mathcal{\chi}_{{\bf a},{\bf
b}} \rightarrow \R$ and then maximizing $\mathcal{S}$ among all
alignments
\begin{eqnarray}
\label{align:maximization} 
\mathsf{S}_0({\bf a},{\bf b}) &= &
\max_{\mathcal{A}} \mathcal{S}(\mathcal{A}; {\bf a},{\bf b}) \notag \\
\mathcal{A}^{\text{opt}} &=& \argmax\, \mathcal{S}(\mathcal{A}; {\bf a},{\bf b}).
\end{eqnarray} 

For the choice of the objective function and its
parameters it is necessary to decide
\begin{enumerate}[(i)]
\item whether we are interested in a locally conserved region or
whether the entire sequences should be considered,
\item how matches and mismatches should be evaluated, and
\item how gaps should influence the overall score oder how a gap
penalty should affect the overall score.
\end{enumerate} 
To address the first issue there are in principle two
types of objective functions available, namely \emph{optimal local
alignment scores} $\Slocal$ and \emph{optimal global alignment scores}
$\Sglobal$.  Optimal global alignment scores involve contributions
from all matches, mismatches and gaps.  Based on this, the optimal
local alignment score is the optimum of all global alignments of all
possible contiguous subsequences of ${\bf a}$ and ${\bf b}$,
\begin{equation}
 \label{eq:local-align-score} \Slocal({\bf a},{\bf b}) =
\max_{\substack{1 \leq i' < i \leq L \\ 1 \leq j' < j \leq M}}
\Sglobal(a_{i'}\ldots a_{i}, b_{j'}\ldots b_{j}).
\end{equation}

The second issue requires the knowledge of a relationship between the
letters of the underlying alphabet.  This is usually realized by
\emph{substitution or score matrices} that assign each pair of letters
a number $\sigma(a,b)$.  Here, we use the most frequently used matrix,
{\tt blosum62} \cite{Heinkoff1992}. In most cases, scoring matrices
are derived from biological data by the scaled log-odd ratio $\Lambda
\frac{P_{a,b}}{f_a f_b}$, where $P_{a,b}$ is the probability of
observing the pair of letters $a$ and $b$, $f_a$ and $f_b$ denote the
background frequencies of observing the letters $a$ and $b$
independently and $\Lambda$ defines a scale \cite{Durbin1998}. The
entries are usually rounded to integers.

Regarding the gaps, one compromises between a computational feasible
and a biological evident penalty function $g$.  That means, each gap
$\Gamma$ of length $l_{\Gamma}$ yields a negative contribution of
$-g(l_{\Gamma})$ to the overall score, which is then defined as
\begin{equation}
\label{eq:alignment-score} \mathcal{S}(\mathcal{A};\,{\bf a},{\bf b})
= \sum_{(i,j) \in \mathcal{A}} \sigma(a_i,b_j) - \sum_{\Gamma}
g(l_{\Gamma})\,,
\end{equation} $g$ is usually a monotonously increasing function of
the gap length.  The alignment algorithms for gaped alignments with
arbitrary gap penalties exhibit a cubic time complexity ($\mathcal{O}(
\max(L,M)^2 \min(L,M) )$).  In practice \emph{affine gap} cost
functions
\begin{equation}
\label{eq:affine-gap} g(l_{\Gamma}) = \open + \ext \, (
l_{\Gamma}-1),\;\;\;\;\; \text{ with } \open \geq \ext > 0
\end{equation} are commonly used, because the computational complexity
then reduces to $\mathcal{O}(LM)$ \cite{Gotoh1982}.  The parameters
$\open$ and $\ext$ are called gap open penalty and gap extension
penalty respectively.  With the above choice one mimics the biological
observations that
\begin{enumerate}[(i)]
  \item longer gaps appear less frequently than shorter ones and
  \item opening a gap is less likely than extending an exisiting one.
\end{enumerate} There is some evidence that this form describes the
natural process of insertions and deletions quite well
\cite{Gonnet1992,Gu1995,Cartwright2006}.

\begin{figure}
\centering 
    \includegraphics[clip,width=0.30\textwidth]{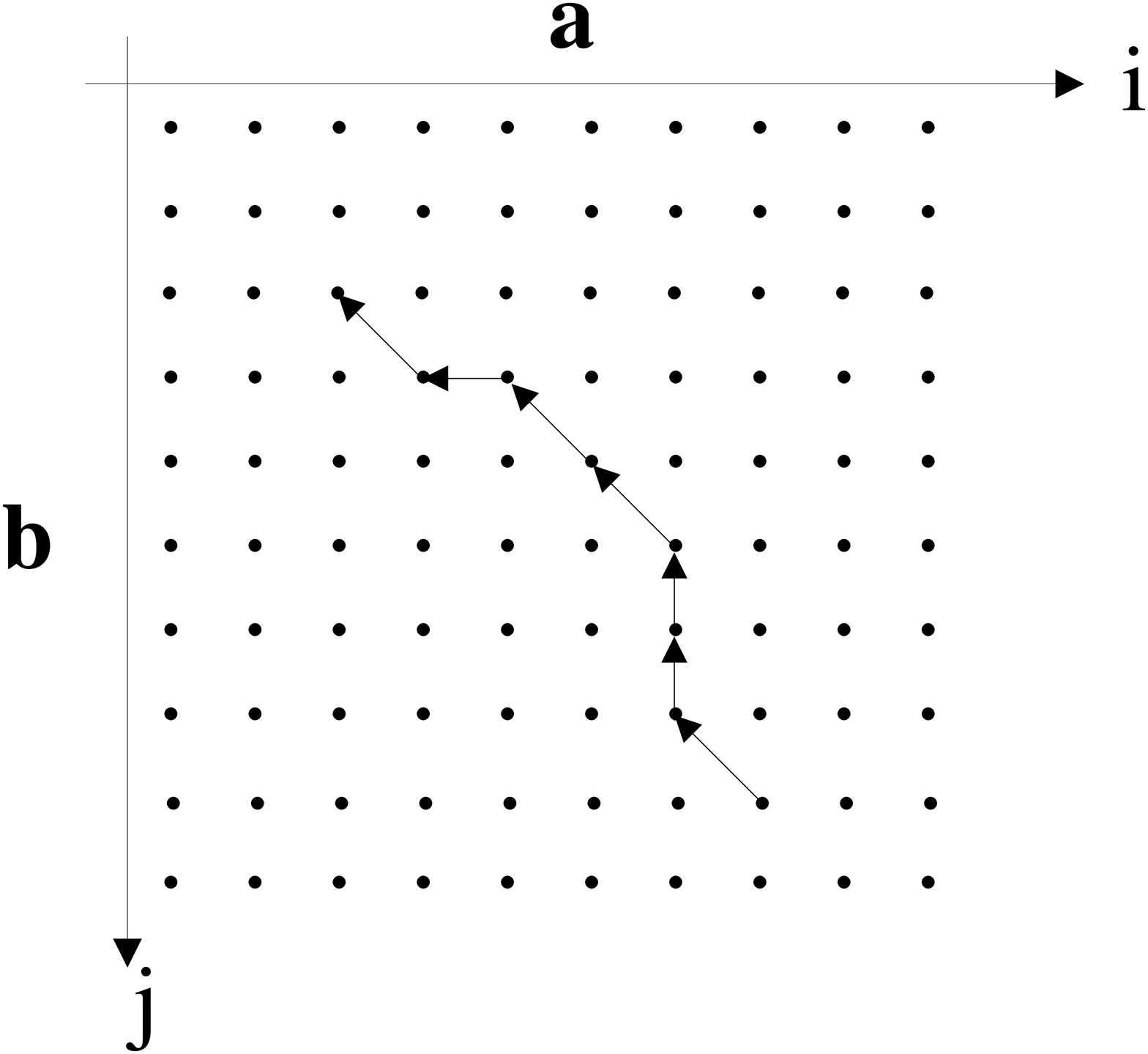}
    \caption{ \label{fig:path} Representation of an alignment as a
directed path in quenched disorder.  The disorder is realized by
random sequences. }
\end{figure} An alignment can be represented as a directed path on a
lattice of size $L \times M$ (see \Fig{fig:path}).  The path is given
by the set of matches and mismatches and gaps in the alignment
$\mathcal{A}$.  Due to the conditions $i_k < i_{k+1}$ and $j_k <
j_{k+1}$ the path is directed.  By convention, we say that each path
element may be orientated in $(-1,-1)$, $(-1,0)$ or $(0,-1)$
direction.  Diagonal elements denote matches or mismatches and
$l_{\Gamma}$ consecutive vertical or horizontal elements correspond to
gaps of length $l_{\Gamma}$ in one of the sequences.

In order to keep the path representation for local alignment unique we
require that
\begin{enumerate}[(i)]
\item the first and the last path element always points in $(-1,-1)$
direction, hence gaps at the begin and end of the alignment never
occur, and
\item a gap in the sequence ${\bf b}$ is not allowed to directly
follow a gap in sequence ${\bf a}$ (see \cite{Muckstein2002}).
\end{enumerate}

The optimal alignment can be computed by the dynamic programming
algorithm (like a transfer matrix method) by Smith and Waterman
\cite{Smith1981}.  For affine gap costs it requires three $(L+1)
\times (M+1)$ matrices $D_{i,j}, P_{i,j}$ and $Q_{i,j}$ that are
computed iteratively \cite{Gotoh1982}.  The matrix element $D_{i,j}$
is the optimal local alignment score of the subsequences $a_{1}\ldots
a_{i}$ and $b_{1}\ldots b_{j}$ given that $a_i$ and $b_j$ are paired.
$P_{i,j}$ and $Q_{i,j}$ are auxiliary matrices storing the optimal
alignment score of the subsequences $a_{1}\ldots a_{i}$ and
$b_{1}\ldots b_{j}$ given that the alignment ends in a gap in either
sequence.  The recursion relation to compute these matrices reads as
\begin{eqnarray}
\label{eq:smithwaterman} D_{i,j} &=& \sigma(a_i,b_j) +
\max \begin{cases} 0\\ D_{i-1,j-1} \\ P_{i-1,j-1} \\ Q_{i-1,j-1} \\
           \end{cases} \notag \\ P_{i,j} &=& \max
	      \begin{cases} D_{i-1,j} - \open \\ Q_{i-1,j} - \open \\
P_{i-1,j} - \ext \\
	      \end{cases} \notag \\ Q_{i,j} &=& \max
	      \begin{cases} D_{i,j-1} - \open \\ Q_{i,j-1} - \ext \\
	      \end{cases}
\end{eqnarray} with the boundary conditions
\[
\begin{array}{lcll} D_{i,0} = P_{i,0} = Q_{i,0} &=& -\infty & \;\;\;
\text{ for } i = 0\ldots L \\ D_{0,j} = P_{0,j} = Q_{0,j} &=& -\infty
& \;\;\; \text{ for } j = 0\ldots M \\
 \end{array}.
\] In a physical interpretation $\sigma(a_i,b_j)$ plays the role of a
random chemical potential with quenched disorder and the gap cost
function $g(l_{\Gamma})$ describes a line tension that forces the
alignment path on a straight diagonal line.

The optimal alignment score, which in physical terms corresponds to
the negative ground-state energy, is given by $\Slocal({\bf a},{\bf
b})= \max\left\{ \max_{i,j}\left\{ D_{i,j} \right\}, 0 \right\}$.  If
$\Slocal({\bf a},{\bf b})=0$, the optimal alignment is the empty
alignment (a path of length $0$). Otherwise the alignment starts at
the position of the maximum of $D_{i,j}$.  The optimal alignment
(ground state) can be determined by a backtrace procedure.  Given that
the current state at position $(i,j)$ is a match or mismatch, the path
is extended in diagonal direction if $D_{i,j} = \sigma(a_i,a_j) +
D_{i-1,j-1}$ and accordingly in vertical or horizontal direction if
$D_{i,j} = \sigma(a_i,a_j) + P_{i-1,j-1}$ or $D_{i,j} =
\sigma(a_i,a_j) + Q_{i-1,j-1}$.  Similar conditions appear, if the
current state is a gap in either sequence.  This is repeated until a
match/mismatch with $D_{i,j} = \sigma(a_i,b_j)$ is met.  In the linear
phase, the ground state might be highly degenerate.  For this reason
one should include additional matrices
$N^{(D)}_{i,j}$,$N^{(P)}_{i,j}$,$N^{(Q)}_{i,j}$ that account for the
degeneracies of the alignments that end at $(i,j)$.  With help of
these matrices it is possible to sample all ground states uniformly.

In the picture of DPRM one usually uses a temporal and a spatial
coordinate which are defined as
\[ t = \frac{1}{2} ( i + j ) \;\;\;\;\text{and} \;\;\;\; x =
\frac{1}{2} ( i-j). \] A local alignment problem is seen as a
dynamical growth process which starts at space-time event $(t_0,x_0)$,
where the dynamic programming matrix $D_{i,j}$ is maximal.  In each
time step described the path is extended by one diagonal, vertical or
horizontal element.  The spatial variable $x-x_0$ describes the
deviation from a straight diagonal line for each time step.  The
stopping condition $D_{i,j} = \sigma(a_i,b_j)$ given that the current
state is a match defines the final point $(t_1,x_1)$ in the
space-time.  We define the roughness of the path as the maximal
deviation from a straight diagonal line, i.e.\ $\Delta = \max_{t_1
\leq t \leq t_0} \left| x(t) - x_0\right|$.  Note that this definition
refers only to the local alignment path and is "time independent" in
contrast to Ref. \cite{Hwa1998}.

Next, we describe the generalization of the alignment problem to a
canonical ensemble of alignments.  Let us consider the canonical
ensemble of all alignments $\mathcal{A}$ for a quenched pair of
sequences.  The partition function at temperature $T$ is given by
\[ Z_T = \sum_{\mathcal{A}} \exp\left[ \mathcal{S}(\mathcal{A};\,{\bf
a},{\bf b})/T \right].
\] This sum can be computed by a generalization of
\Eq{eq:smithwaterman} \cite{Miyazawa1995},
\begin{eqnarray}
\label{eq:smithwaterman-partfunc} Z^D_{i,j} &=& \left( 1 +
Z^{D}_{i-1,j-1} + Z^{P}_{i-1,j-1} + Z^{Q}_{i-1,j-1} \right) \cdot
e^{\sigma(a_i,b_j) / T } \notag \\ Z^P_{i,j} &=& \left( Z^D_{i-1,j} +
Z^Q_{i-1,j} \right) \cdot e^{-\open / T} + Z^P_{i-1,j} \cdot e^{-\ext
/ T} \notag \\ Z^Q_{i,j} &=& Z^D_{i,j-1} \cdot e^{-\open / T } +
Z^Q_{i,j-1} \cdot e^{- \ext / T }
\end{eqnarray} with the boundary conditions
\[
\begin{array}{lcll} Z^D_{i,0} = Z^P_{i,0} = Z^Q_{i,0} &=& 0 & \;\;\;
\text{ for } i = 0\ldots L \\ Z^D_{0,j} = Z^P_{0,j} = Z^Q_{0,j} &=& 0
& \;\;\; \text{ for } j = 0\ldots M.
\end{array}
\] Since an alignment may start anywhere and may also include the
empty alignment, the full partition function is given by
\[ Z = 1 + \sum_{i=1}^{L} \sum_{j=1}^{M} Z^D{i,j}.
\] Note that contributions from $Z^P$ and $Z^Q$ are explicitly
excluded because they are auxiliary only and contain non-canonical
alignments.  In the limit $T \rightarrow 0$
\Eq{eq:smithwaterman-partfunc} reduces to the recursion relation of
the original Smith-Waterman algorithm \Eq{eq:smithwaterman}.  Once the
transfer matrices $Z^D_{i,j}$, $Z^P_{i,j}$ and $Z^Q_{i,j}$ are
determined it is possible to directly draw alignments from the
canonical distribution $P_T\left[\mathcal{A}\right] = \exp\left[
\mathcal{S}(\mathcal{A};\,{\bf a},{\bf b})/T \right] / Z_T$ with zero
autocorrelation.  This direct Monte Carlo algorithm was proposed by
M\"uckstein et.\ al.\ \cite{Muckstein2002} for local alignment.  A
general description of such methods are presented in the textbook of
Durbin et.\ al. \cite{Durbin1998}.

%
%
\section{Results}
\label{sect:results} To study properties of the linear-logarithmic
phase transition, we generated ensembles of $n^{\text{sample}}$ random
sequences which were drawn from the distribution $P({\bf
a})=\prod_{i=1}^{L} f_{a_i}$, where $f$ are the amino acid frequencies
that were derived together with {\tt blosum62} \cite{Heinkoff1992}.
Furthermore we only consider pairs of sequences of equal length, i.e.\
$L=M$, between $L=40$ and $L=5120$.  It turned out that for the
finite-size-scaling analysis, that is discussed in the following, only
system sizes with $L \geq 480$ yield consistent results.  The number
$n^{\text{sample}}$ of samples varied between $6400$ for $L=480$ and
$800$ for the largest system.
 
For each sample, $n^{\text{align}} = 100$ alignments were drawn from
the canonical ensemble at various temperatures $T$ using the
backtracing procedure as described above.  We used different gap-open
parameters $\open$ and temperatures $T$ between $T=0$ and $T=4$.  The
gap extension parameter $\ext$ was set to $1$ throughout.

The case $T=0$ corresponds to optimal alignments (ground states).
Note that for small gap costs (i.e.\ in the linear phase) the
ground-state degeneracy grows exponentially with the system size,
whereas in the logarithmic phase usually only a few optimal alignments
are observed.  Thermal averages and averages over ground states over a
fixed realization of a sequence pairs will be denoted as $\langle
\cdot \rangle_T$ and $\langle \cdot \rangle_0$, respectively.
Averages over realizations of random sequence pairs will be written as
$[ \cdot ]$ in the following.  In statistical mechanics of disordered
systems the latter one is often called average over the disorder.

So as to describe the linear-logarithmic transition, we considered
different observables described in the subsequent sections.

%
%
\subsection{Geometric properties}
Here, we describe  the results for the number $N_m$ of paired letters 
(matches  plus mismatches).
This quantity turned out to be an adequate quantity to extract properties of
the phase transition, such as the critical gap costs $\alpha_c(T)$ and
the scaling behavior close to criticality.

We consider the averaged  number $\left[ \langle N_m \rangle_T \right]
/ L$ of paired letters per sequence length as a function of gap opening 
penalty $\open$ and temperature  $T$.  Hence, it is the  fraction of 
matches/mismatches with  respect  to  the  maximal  possible  number  of  
pairs.   
This observable corresponds to the percolation probability (the probability
that  a  geometrical object  spans  the  entire  lattice) in  standard
percolation theory.  Usually  a crossover from $1$ to  $0$ is observed
when passing the phase  boundary.  Here, a perfectly percolating local
alignment $N_m  = L$  is hardly found  even in the  percolating phase.
Nevertheless, we  applied the  same finite-size-scaling  analysis as
for  the usual  percolation probability,  because $\left[  \langle N_m
\rangle_T \right] /  L$ is in the order of unity  in the linear regime
and vanishes in the logarithmic regime.

\begin{figure}
  \centering 
    \includegraphics[clip,width=0.48\textwidth]{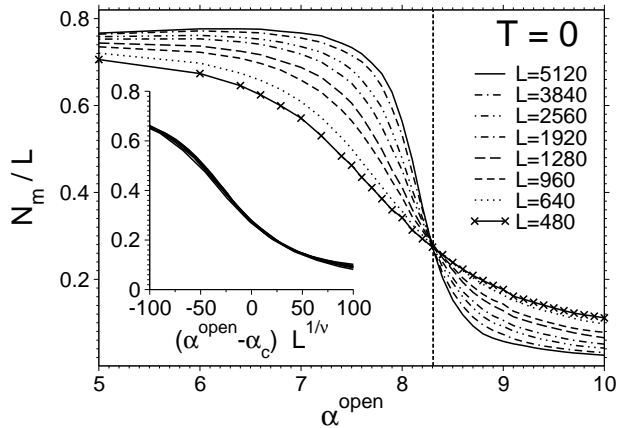}
    \caption{  \label{fig:FSSMatch}  Results for the  number $N_m$  
     of  aligned  letters
     (matches  plus mismatches)  as  a function  of gap opening  penalty
     $\open$. Curves for  different sequence lengths intersect at
     the  critical parameter  $\alpha_c$.  
     For a more clear presentation, single data points are only  shown for one system size.  
     Inset:
     After  rescaling   the  abscissa  with   appropriate  values  for
     $\alpha_c$ and $\nu$ the data listed in Tab. \ref{tab:expo-match} 
     collapses on a single master curve.
     }
\end{figure}
\Fig{fig:FSSMatch} displays $\left[ \langle N_m \rangle_T \right] / L$
as a function of the gap open penalty $\open$ for different lengths
$L$ and  zero temperature.  The curves for  different sequence lengths
intersect at  the critical value  $\alpha_c$ as expected for  a second
order   phase   transition.    Using   finite-size-scaling   theory
\cite{Stauffer}, we may extrapolate  data from finite sequence lengths
to the thermodynamic limit $L  \rightarrow \infty$.  In this limit the
observable $\left[ \langle N_m \rangle_T \right] / L $ is in the order
of $1$ below the threshold, i.e.\ $\open < \alpha_c$, and it jumps to
$0$  exactly at  $\alpha_c$.  In  finite  systems, $L  < \infty$,  the
crossover extends over a range $\sim L^{-1/\nu}$ 
as can  be seen in \Fig{fig:FSSMatch}.  Scaling
theory  leads us to expect that  the  behavior of  
$\left[  \langle N_m  \rangle_T \right](\open; T; L)$ close to 
criticality is described by
\begin{equation}
\label{eq:finite-size-scaling}
\left[ \langle N_m \rangle_T \right](\open; T; L)/L = 
 f\left( (\open - \alpha_c) L^{1/\nu} \right),
\end{equation}
where   $f$  is   an   universal  scaling   function and the exponent 
$\nu$ describes the divergence of the "correlation length" 
at the critical point $\open = \alpha_c$.  
We   used
\Eq{eq:finite-size-scaling}  to extract numerical values for
the critical  exponents $\nu$
and  the critical  gap  costs $\alpha_c$ with high precession from  all data  
for a  fixed temperature  simultaneously.  The  fit  is performed  by minimizing  a
weighted-$\chi^2$-like  objective  function  $Q(\alpha_c,\nu)$  \cite{Houdayer2004},
that measures the  distance (in units of the standard error)  of the data
from the  (a prior  unknown) master curve.   For the example in \Fig{fig:FSSMatch},  
i.e.\ $T=0$,  we obtained  $\alpha_c=8.306(4)$ and $\nu=1.58(5)$ with acceptable
quality   of  $\hat{Q} \equiv Q(\alpha_c,\nu) = 2.2$   
($\hat{Q}$   should   be  in   the   order  of   $1$
\cite{Houdayer2004}).  Statistical  errors  have  been  determined  by
bootstrapping  \cite{Efron1979,practicalGuide2009}.    Repeating  this  analysis   for  several
temperatures  one  may  probe  the critical  line  $\alpha_c(T)$  (see
below).  

We also tested a related quantity, which is defined as
the number  of matches  / mismatches  plus the number  of gaped letters  
$N_m + N_g$, which results in the  same critical exponent and critical point within error
bars (not shown). Gaps seem to play only a  marginal role  close to the  critical point.
We observe that the roughness  $\Delta$ of  the alignment  path as defined above 
at  the critical point diverges only logarithmically  with the sequence length.
This supports the equivalence the description of the description of both variants of 
the this quantity.

\begin{figure}
  \centering 
   \includegraphics[clip,width=0.48\textwidth]{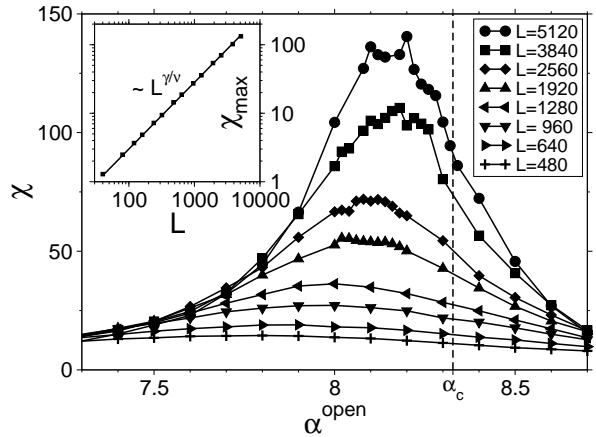}
    \caption{  \label{fig:Susc} Critical  fluctuations  of $N_m$.  The
     positions of the peaks approach the critical value $\alpha_c$ the
     and their heights diverge like $L^{\gamma / \nu}$.  Inset: fit of
     $\chi_{\max}(L)$ to the scaling form $\sim L^{\gamma / \nu}$.  }
\end{figure}
Next,  we  study  the   critical  fluctuations of $N_m$.  
We define the susceptibility-like quantity 
$\chi = (\left[ \langle N_m^2 \rangle_T \right] - \left[  \langle N_m \rangle_T \right]^2)/L$ as a
function  of $\open$  (see  \Fig{fig:Susc}).

Close  to the  critical
point $\chi$ diverges like $\chi  \sim L^{\gamma / \nu}$.  In order to
extract  the height  of the  maxima from  $\chi(\open)$  we performed
parabolic fits in  the form $\chi(\open)=-C ( \alpha_c(L)-\open)^2
+ \chi_{\max}(L)$ for each system size $L$.  The exponent $\gamma$ itself
is determined by  a fit of $\chi_{\max}(L)$ to  the scaling form $\sim
L^{\gamma / \nu }$.  For $T=0$, we obtain $\gamma / \nu = 0.95(1)$.

One  may  also use  the  scaling  of  $\alpha_c(L)$ to  determine  the
critical value $\alpha_c$  and the critical exponent $\nu$  as a cross
check via $\alpha_0(L) = \alpha_c - A L^{-\nu}$.  When restricting the
range to $L  \geq 1280$ we obtain $\nu=1.4(3)$  for $T=0$ which agrees
within the error  bars with the value stated above. 
Furthermore the critical value $\alpha_c$ agrees  within 1.5 standard
deviations.  We also checked that those values agree for FTA, $T=1$.

\begin{figure}
  \centering \includegraphics[clip,width=0.48\textwidth]{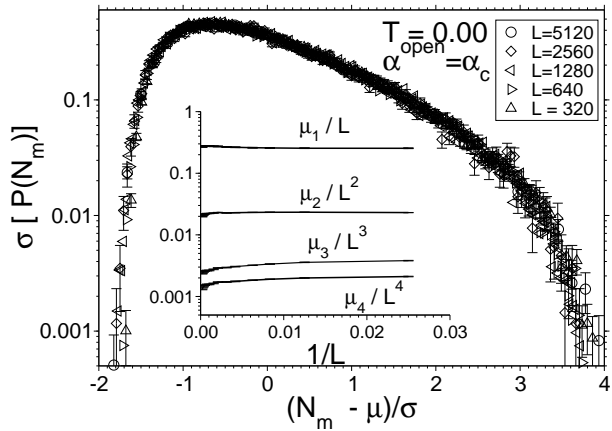}
    \caption{  \label{fig:DistrMatch}  Rescaled  distributions of  the
    observable  $N_m$  for  $T=0$.   Inset: scaling  analysis  of  the
    moments. The first  moment scales as $\mu_1 \sim  L$. Higher order
    moments increase  slower than $\mu_k \sim  L^k$.   }
\end{figure}

Alternatively, one  can determine $\gamma$ from  the second moment
of  the averaged distribution  $[P(N_m)]$ of $N_m$ at $\open=\alpha_c$.  
Hence,  we performed 
further simulations at criticality  with a larger sample size (for
the largest system size, $n^{\text{sample}} \approx 2 \times 10^4$ for
$T=0$  and $n^{\text{sample}}=1.6  \times 10^4$  for $T  >  0$).  This
allowed us  to cross check the  value of $\gamma$  and  to   extract  
higher  moments  of  the  distribution.

\Fig{fig:DistrMatch} displays distributions $[P(\langle N_m \rangle_T)]$
for  $T=0$  (similar results  have  been obtained  for  $T  > 0$).  The
distributions have been  rescaled to zero mean and  unit variance.  In
all cases  we observe that the  first moments $\mu_1$  scale as $\mu_1
\sim L^{1+\epsilon}$  with $\epsilon =  0$ within the  errorbars.  
For the second moment one would  expect the scaling behavior 
$\mu_2 \equiv L \chi \sim L^{1+\gamma/\nu}$.  
Indeed we find $\gamma/\nu  = 0.955(8)$ for OA 
by a  least square fit. This is consistent with the numerical value 
obtained by the height of maxima.
The third and the fourth  moment scale as $\mu_3 \sim L^{2+\gamma_3}$
and $\mu_4  \sim L^{3+\gamma_4}$ respectively.  For  temperatures 
$T < 2$  both   exponents  $\gamma_3$  and  $\gamma_4$   agree  within  the
statistical   errors.   

\begin{figure}
  \centering 
\includegraphics[clip,width=0.48\textwidth]{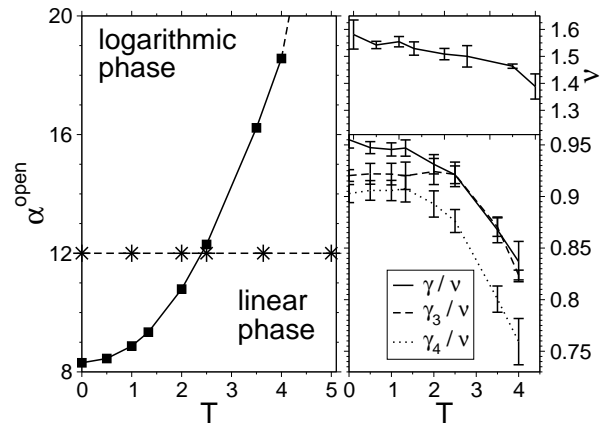}
    \caption{ \label{fig:PHASE} 
    Results for FTA. Left: Phase diagram for FTA. The linear phase is located below the critical line.
    Right: critical exponents as a function of the temperature. 
    }
\end{figure}

The resulting critical values $\alpha_c$ and critical exponents $\nu$,
$\gamma$,      $\gamma_3$,$\gamma_4$       are      summarized      in
\Tab{tab:expo-match}.  
All ratio of exponents, $\gamma / \nu$,  $\gamma_3/ \nu$   and
$\gamma_4 / \nu$, are  in the  order of  $1$ and seem  to increase  with the
temperature. Note that for a perfectly  one-parameter scaling of the complete distribution
with $\mu_1 \sim L$ one would expect $\gamma/\nu = \gamma_3/\nu = \gamma_4/\nu = \ldots = 1$.
This property is only approximately fulfilled according to our data. 
This is shown in \Fig{fig:PHASE}, where also the resulting  phase diagram  is displayed.

The standard  order parameter in  percolation problems is the relative 
size of the  largest cluster. 
Since local sequence alignments (and its interpretation as DPRM) 
exhibits one spacial dimension, the observable $N_m / L$ can also be 
interpreted as 
order parameter, which is one if the alignment covers the entire sequences.
The usual finite-size-scaling ansatz for the relative size of the 
largest cluster order parameter reads as 
\begin{equation}
\label{eq:fss-cluster}
\left[ \langle N_m \rangle_T \right](\open; T; L)/L  =
L^{-\beta^{\text{geo}}/\nu} f  \left( (\open  -\alpha_c) L^{1/\nu} \right),
\end{equation}
where the exponent $\beta^{\text{geo}}$ describes the divergence of the largest cluster.
By comparing this relation with \Eq{eq:finite-size-scaling} we may infer $\beta^\text{geo}=0$ and 
verify that the scaling relation $\gamma + 2\beta^\text{geo} = d \nu$ \cite{Stauffer} (with $d=1$ in our case) is again only 
approximately fulfilled. 
We confirmed that $\beta = 0$ within the errorbars by considering $\beta$ as free parameter in  the finite-size-scaling analysis for $N_m$. 

As mentioned above, the roughness only grows logarithmically with the system size.
This implies that the fractal dimension $d_r$ of the alignment path equals the topological dimension $d=1$, which is in agreement with the scaling relation $d_r = d -\beta^\text{geo}/\nu$ in a trivial way.

\begin{table}
\begin{center}
  \begin{tabular}{r@{.}l|r@{.}l|r@{.}l|r@{.}l|r@{.}l|r@{.}l}
    \hline
    \multicolumn{2}{c|}{$T$}&
    \multicolumn{2}{c|}{$\alpha_c$}&
    \multicolumn{2}{c|}{$\nu$}&
    \multicolumn{2}{c|}{$\gamma / \nu$}&
    \multicolumn{2}{c|}{$\gamma_3 / \nu$}&
    \multicolumn{2}{c}{$\gamma_4 / \nu $}\\
    \hline
    $0$&$00$  & $8$&  $306(4)$	& $1$& $58(5)$   &$0$& $955(8)$	 &$0$&$920(5)$  & $0$&$903(9)$\\
    $0$&$50$  & $8$&  $450(2)$	& $1$& $54(1)$   &$0$& $947(6)$	 &$0$&$92(1)$   & $0$&$91(1)$\\
    $1$&$00$  & $8$&  $871(4)$	& $1$& $55(2)$   &$0$& $946(6)$	 &$0$&$92(1)$   & $0$&$91(1)$\\
    $1$&$33$  & $9$&  $339(3)$	& $1$& $53(2)$   &$0$& $948(8)$	 &$0$&$92(1)$   & $0$&$91(1)$\\
    $2$&$00$  & $10$& $791(3)$	& $1$& $51(2)$   &$0$& $931(9)$	 &$0$&$92(1)$   & $0$&$89(1)$\\
    $2$&$50$  & $12$& $296(4)$	& $1$& $50(4)$   &$0$& $921(9)$	 &$0$&$92(1)$   & $0$&$88(1)$\\
    $3$&$50$  & $16$& $227(1)$	& $1$& $46(1)$   &$0$& $87(1)$	 &$0$&$870(8)$  & $0$&$80(1)$\\
    $4$&$00$  & $18$& $557(2)$	& $1$& $38(5)$   &$0$& $84(2)$	 &$0$&$924(5)$  & $0$&$76(2)$\\

    \hline
  \end{tabular}
 \caption{
   \label{tab:expo-match}
   Critical  gap open penalty  $\alpha_c$ and  critical  exponents $\nu$,
   $\gamma$, $\gamma_3$,$\gamma_4$ for the observable $N_m$.  }
\end{center}
\end{table}

%
%
\subsection{Energetic properties}
\label{sec:energetic-properties}
As mentioned above, the size of the largest cluster is usually 
regarded as the order parameter in  percolation problems.
In the non-percolating phase, the size of the largest cluster typically grows 
logarithmically with the system size whereas in the percolating phase its extension 
is comparable to the system size \cite{Stauffer}.
The average  score of
local  alignments  exhibits  the   same  crossover  when  crossing  the
linear-logarithmic boundary.  
For this  reason we regard the average
score $\left[ S \right]/L$ (OA), or free energy $\left[ F_T \right]/L$ (FTA), 
per length as a second order  parameter, as in \cite{Sardiu2005}.  
Note that there is  no direct geometrical interpretation for this quantity.

\begin{figure}
  \centering 
    \includegraphics[clip,width=0.48\textwidth]{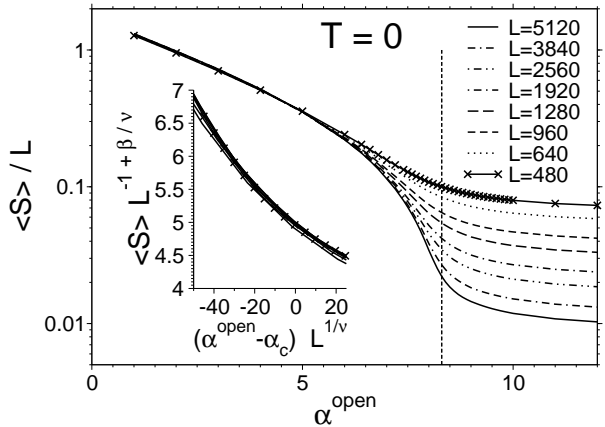}
    \caption{ \label{fig:fss-score}  Finite-size-scaling  analysis for
    the average score per length.  }
\end{figure}

Scaling theory states that the order parameter scales as
\begin{equation}
\label{eq:fss-score}
 \left[  F_T \right]/L  =  L^{-\beta/\nu} f  \left( (\alpha  -
 \alpha_c) L^{1/\nu} \right)
\end{equation}
with some universal scaling  function $f$.  This allows one to
extract the critical value  $\alpha_c$ and exponents $\nu$ and $\beta$
from the data with the same method as described above.  Here, we fixed
$\alpha_c$ and $\nu$ with the  values that have been obtained from the
data collapse  for the  observable $N_m$ and  regard $\beta/\nu$  as a
free  parameter.   The result  for  $T=0$ is  shown  in
\Fig{fig:fss-score}. The quality  of  these fits  varied
between $\hat{Q}=1.58$ for $T=1.00$ and $\hat{Q}=7.49$ for $T=4.00$.

By regarding  $\nu$ and $\beta$ as  free parameters, we  also used the
scaling  form  \Eq{eq:fss-score}  as  a  second cross  check  for  the
exponent  $\nu$.  Within  the error  bars  it is  comparable with  the
results of the  finite-size-scaling analysis for  the observable $N_m$
(for example  $\nu=1.50(7)$ for  $T=0$).  For larger  temperatures ($T
\geq 2$) only system sizes $L \geq 1280$ led to convincing results
for this kind of check.

\begin{table}[t!]
\begin{center}
  \begin{tabular}{r@{.}l|r@{.}l|r@{.}l} 
    \hline
    \multicolumn{2}{c|}{$T$}&
    \multicolumn{2}{c|}{$\beta/\nu$}&
    \multicolumn{2}{c}{$\omega$}\\
    \hline
    $0$&$00$  & $0$&  $6324(07)$	& $0$& $386(5)$ \\ 
    $0$&$50$  & $0$&  $6406(10)$	& $0$& $391(5)$ \\ 
    $1$&$00$  & $0$&  $6443(11)$	& $0$& $387(6)$ \\ 
    $1$&$33$  & $0$&  $6564(09)$	& $0$& $380(9)$ \\ 
    $2$&$00$  & $0$&  $6835(10)$	& $0$& $36(1)$  \\ 
    $2$&$50$  & $0$&  $7081(08)$	& $0$& $33(1)$  \\ 
    $3$&$50$  & $0$&  $7691(06)$	& $0$& $23(2)$  \\ 
    $4$&$00$  & $0$&  $7958(05)$	& $0$& $20(2)$  \\ 
    \hline
  \end{tabular}
 \caption{
   \label{tab:expo-score}
   Critical exponents for the average  score / free energy per length.
   $\beta/\nu$   was   obtained   from   finite-size   scaling   (see
   \Fig{fig:fss-score}) and cross checked via the scaling of the first
   moment of the  score distribution at criticality $\alpha=\alpha_c$.
   The  exponent  $\omega$ describe the 
   fluctuations of the score distribution at $\alpha=\alpha_c$.  }
\end{center}
\end{table}
As can be  seen in the second column  of table \Tab{tab:expo-score}, the
free energy  per length $\left[  F_T \right]/L$ decreases  like $\sim
L^{-\beta / \nu}$, where $\beta / \nu$ increases monotonously with the
temperature from $0.6324(7)$ to $0.7958(5)$. For small temperatures, i.e.
$T  <  2$,   the  exponent  $\beta  \approx  1$   is  not  temperature
dependent. Hence the phase  behavior regarding the exponent $\beta$ is
not universal  any more when exceeding $T=2$.

\begin{figure}
  \centering
  \includegraphics[clip,width=0.48\textwidth]{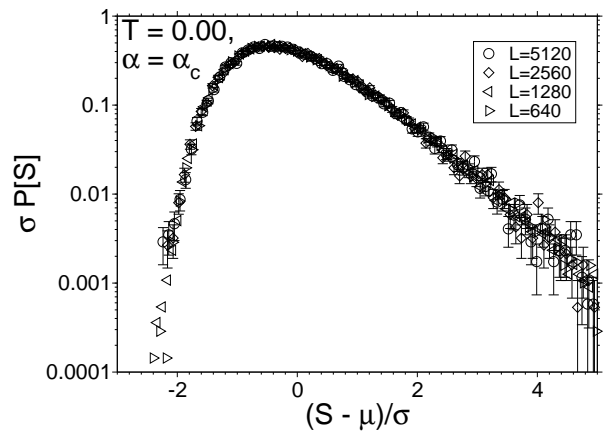}
    \caption{  \label{fig:DistrScore}   Rescaled  score  distributions
    $T=0$. The free-energy distributions of FTA look comparable.  }
\end{figure}

%
%
\subsection{Free-energy distributions close to criticality}
In   analogy  to  the   distribution  of   the  observable $N_m$  in
\Fig{fig:DistrMatch}, the resulting rescaled score distributions right
at $\open = \alpha_c$  are shown in \Fig{fig:DistrScore}. 
Simulations for FTA yield comparable results.
We performed an  analysis of the  moments of $P[S]$ and $P[F]$ respectively.
The   first  moment   scales  as   $\mu_1  \sim
L^{1-\beta/\nu}$  as expected from the finite-size analysis shown in 
\Fig{fig:fss-score}. We  checked that  the fit  parameters
agree  with   those  from  finite-size   scaling.   

Regarding the second moment, no
divergence was observed.  Its scaling behavior is
given by $\mu_2  \sim L^{1-\omega}$ with $\omega >  0$.    
The  resulting fit parameters are  listed in  \Tab{tab:expo-score}.  
In the  limit $\open \rightarrow \infty$ and $L  \rightarrow \infty$,
 the score distribution
is predicted to follow a Gumbel distribution
\begin{equation}
\label{eq:gumbel}
P_{\text{gumbel}}(S) = 
\lambda \exp\left[ -\lambda (S-S_0) -\exp(-\lambda ( S - S_0) ) \right]
\end{equation}
(according to the Karlin-Altschul-Dembo theory \cite{Karlin1990,Karlin1992,Dembo1994}).

In the  linear phase the conditions  of this theory are  not valid any
more.  Interestingly,  right at  the critical point,  the shape  of the
distributions are well described by a Gumbel distribution, at least in
the  high  probability  region  (down  to $P[S]  \sim  10^{-4}$).   In
previous studies  we observed parabolic  corrections to \Eq{eq:gumbel}
that occur  in the  far right tail  of the optimal  score distribution
\cite{Hartmann2002,Wolfsheimer2007}.   The  corrected distribution  is
empirically well described by
\begin{equation}
\label{eq:mod-gumbel}
P(S)      =      \frac{1}{z'}     P_{\text{gumbel}}(S)      \exp\left[
-\lambda_2(S-S_0)^2 \right],
\end{equation}
where  $\lambda_2$ is  a  correction parameter  and the  normalization
constant $z'$  is indistinguishable from $1$.  We  found evidence that
$\lambda_2$ vanishes for $L  \rightarrow \infty$ but persists even for
gapless alignment $\open = \infty$ \cite{Wolfsheimer2007}.

Here, we  extend this study to finite-temperature  alignment, i.e.\ to
the  free-energy  distribution  as   a  generalization  of  the  score
distribution.    We    employed   generalized   ensemble   Monte-Carlo
simulations  combined  with  Wang-Landau  sampling \cite{Wang2001}  in  the
sequence     space     (details     can     be     found     elsewhere
\cite{Wolfsheimer2008a}).  The (production) run for $L=120$ employed $4.8
\times 10^{7}$ Monte Carlo steps for each distribution over the disorder.

In the following, we use  the  phase  diagram  as  a guide  to  study  
the  free-energy distribution  for various  temperatures. We  kept the  gap-costs fixed
($\open=12$,  $\ext=1$)  and only  varied  the temperature  (between
$T=0$  and $T=5$).  The  interpolating points are  indicated by  stars in  the phase
diagram in \Fig{fig:PHASE}.

\begin{figure}
\begin{center}
 \includegraphics[clip,width=0.48\textwidth]{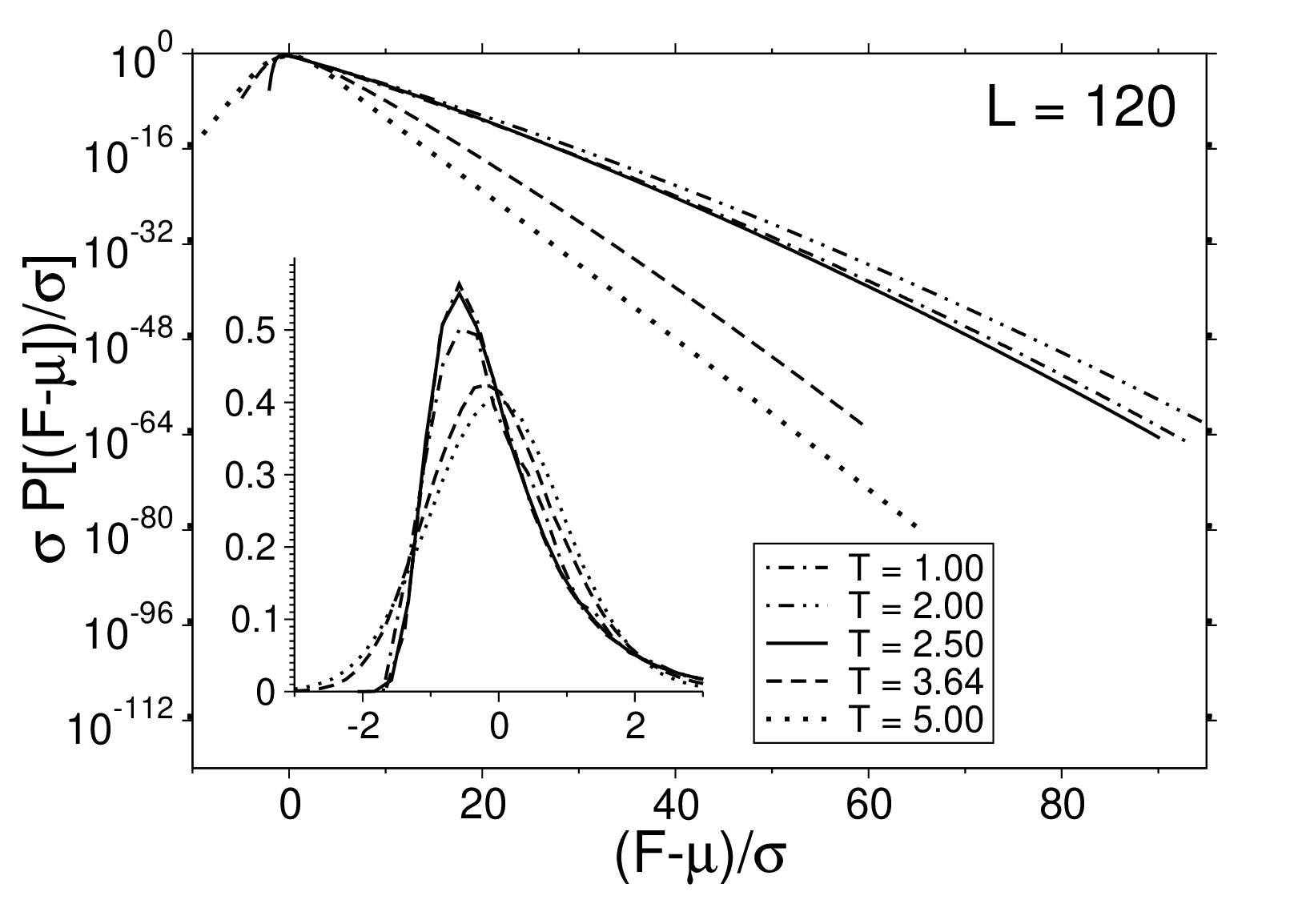}
\end{center}
{
  \caption{\label{fig:finite-temper-distr}     Rescaled    free-energy
  distribution of finite-temperature alignments. At $T=2.50$ and below,
  the data  is well described  by a modified Gumbel  distribution. For
  large  temperatures an exponential  tail is  observed.\newline Inset:
  The same data shown with  a linear ordinate. In the high probability
  region  the  data for  $T=5.00$  is  well  described by  a  Gaussian
  distribution.  } }
\end{figure}

\begin{table}
\begin{center}
  \begin{tabular}{r@{.}l|r@{.}l|r@{.}l|r@{.}l}
    \hline
    \multicolumn{2}{c|}{$T$}&
    \multicolumn{2}{c|}{$\lambda$}&
    \multicolumn{2}{c|}{$10^4 \lambda_2$}&
    \multicolumn{2}{c}{$s_0$}\\
    \hline
    $0$&$00$  & $0$&$2966(4)$	& $3$&$182(1)$  &$37$&$4(1)$	\\
    $1$&$00$  & $0$&$2924(1)$	& $2$&$900(5)$  &$24$&$6(1)$	\\
    $2$&$00$  & $0$&$2907(2)$	& $3$&$122(7)$  &$31$&$56(6)$	\\
    $2$&$50$  & $0$&$2980(2)$	& $3$&$16(1)$   &$38$&$29(7)$	\\
    \hline
  \end{tabular}
 \caption{
   \label{tab:finite-temperature-fit}
   Fit   parameters  of   least  $\chi^2$-fits   of   the  free-energy
   distributions     to    the     modified     Gumbel    distribution
   \Eq{eq:mod-gumbel} for $L=120$ in the logarithmic phase.  }
\end{center}
\end{table}

In the logarithmic regime ($T=0,1,2,2.5$) the free-energy distribution
is    well   described   by    the   modified    Gumbel   distribution
\Eq{eq:mod-gumbel} (see  \Fig{fig:finite-temper-distr}).  Note that we
have again rescaled the distributions  to unit variance and zero mean.
The  fit parameters  only change  slightly with  the  temperature (see
\Tab{tab:finite-temperature-fit}).

The crossover  from the logarithmic  to the linear regime  comes along
with  a change  of  the  skewness, as  can  be seen  in  the inset  of
\Fig{fig:finite-temper-distr}.  In the high probability region, 
for a large value $T=5.00$,  a Gaussian distribution  describes the  data well.  This was
confirmed  by  a Kolmogorov-Smirnov  test  that  yielded  a p-value  of
$0.14$.   For $T=1/0.275  \approx 3.64$  the evidence  for  a Gaussian
distribution is much  smaller (a p-value of $2  \times 10^{-11}$).  We
also checked  that the change of  the shape is accompanied by  a change from
logarithmic to linear growth of typical free energies (the position of
the maximum) with  the sequence length, i.e.\ the  free energy becomes
extensive (not shown here).  This result can be understood
in the following way: The partition  functions that  appears  in the  transfer matrix  calculation
\Eq{eq:smithwaterman-partfunc} become  (more or less)  independent and
hence  factorize in the linear phase.  
The  total free  energy  decomposes into  a sum  of
independent contributions and the central limit theorem applies.

When  considering  the rare-event  tail  at  higher temperatures,  the
free-energy distribution  is rather exponential than  Gaussian, as can
be seen in the  main plot of \Fig{fig:finite-temper-distr}.  Hence, we
observe  a  crossover  from   a  Gaussian  distribution  in  the  high
probability  region  to the  characteristic  exponential  tail of  the
Gumbel distribution.   With the same argumentation as  for the optimal
alignment,  sequence   pairs  appearing  in  the   tail  feature  high
similarities.  The overall free energy  is dominated by the ground
state. This  was confirmed  by looking at  the difference  between the
free energy and the ground-state energy for those sequences that occur
in the tail of the distribution.  The summation in the transfer matrix
are virtually replaced by  maximization yielding an exponential tail.
The finite-size  effect, that is  responsible for the curvature  of the
optimal alignment  statistics seems to be of  marginal order  in this
case.

\section{Summary}
\label{sect:discussion}
We presented  
a finite-size-scaling analysis  of the linear-logarithmic phase
transition of finite-temperature protein
sequence alignment. 
This phase transition is crucial to determine the set of
parameters where the alignment is of highest sensitivity.
We have used the {\tt blosum62} scoring matrix together with affine gap costs,
which is the most frequently used scoring system for actual
 database queries. This goes much beyond previous studies, 
which have investigated only simple scoring systems.

  Two order parameters were studied  in detail: the
number of  matches (i.e.\ the  alignment length) and the  average free
energy per length. We have analyzed the 
phase transition using finite-size scaling techniques. Using sophisticated
algorithms, large systems could be studied, such that corrections
to finite-size scaling are negligible.
The resulting critical line $a_c(T)$ in the range $T=0 \ldots
4$  provides a  guide  for biological  applications where  suboptimal
alignments play an important role.

Numerical values of the critical exponents $\nu$, $\gamma$ and $\beta$
suggest that the percolation transition is not universal with respect
to different temperature values.  

The free-energy distribution, which can  be seen as a generalization of
the  score distribution  over  random sequences,  crosses  over from  a
modified Gumbel  distribution with a parabolic correction  in the tail
given  by \Eq{eq:mod-gumbel} in  the logarithmic  phase to  a modified
Gaussian  distribution with  a linear  rare-event tail  in  the linear
phase. This is another example showing that the 
large-deviation properties of
 order-parameter distributions change significantly close to
phase transitions.

\section*{Acknowledgments}
 This project was  supported  by  the  German {\em  VolkswagenStiftung}
  (program  ``Nachwuchsgruppen   an  Universit\"aten'')  and   by  the
 European  Community DYGLAGEMEM program.   
 SW  was partially supported by the University Paris Descartes.
 The simulations  were performed at the GOLEM I cluster for scientific computing at the University of Oldenburg (Germany). 

\bibliography{temperalign}

\end{document}